\documentclass[12pt,epsf,amssymb]{article}

\usepackage{graphicx}
\usepackage{axodraw}
\usepackage{amsfonts}
\usepackage[usenames]{color}
\usepackage{amsmath}

\begin{document}
\baselineskip 0.6cm
\newcommand{\gsim}{ \mathop{}_{\textstyle \sim}^{\textstyle >} }
\newcommand{\lsim}{ \mathop{}_{\textstyle \sim}^{\textstyle <} }
\newcommand{\vev}[1]{ \left\langle {#1} \right\rangle }
\newcommand{\bra}[1]{ \langle {#1} | }
\newcommand{\ket}[1]{ | {#1} \rangle }
\newcommand{\Dsl}{\mbox{\ooalign{\hfil/\hfil\crcr$D$}}}
\newcommand{\nequiv}{\mbox{\ooalign{\hfil/\hfil\crcr$\equiv$}}}
\newcommand{\nsupset}{\mbox{\ooalign{\hfil/\hfil\crcr$\supset$}}}
\newcommand{\nni}{\mbox{\ooalign{\hfil/\hfil\crcr$\ni$}}}
\newcommand{\EV}{ {\rm eV} }
\newcommand{\KEV}{ {\rm keV} }
\newcommand{\MEV}{ {\rm MeV} }
\newcommand{\GEV}{ {\rm GeV} }
\newcommand{\TEV}{ {\rm TeV} }

\def\diag{\mathop{\rm diag}\nolimits}
\def\tr{\mathop{\rm tr}}

\def\Spin{\mathop{\rm Spin}}
\def\SO{\mathop{\rm SO}}
\def\O{\mathop{\rm O}}
\def\SU{\mathop{\rm SU}}
\def\U{\mathop{\rm U}}
\def\Sp{\mathop{\rm Sp}}
\def\SL{\mathop{\rm SL}}
\def\simgt{\mathrel{\lower2.5pt\vbox{\lineskip=0pt\baselineskip=0pt
           \hbox{$>$}\hbox{$\sim$}}}}
\def\simlt{\mathrel{\lower2.5pt\vbox{\lineskip=0pt\baselineskip=0pt
           \hbox{$<$}\hbox{$\sim$}}}}

\def\change#1#2{{\color{blue} #1}{\color{red} [#2]}\color{black}\hbox{}}


\begin{titlepage}

\begin{flushright}
UCB-PTH-05/41 \\
LBNL-59212 \\
\end{flushright}

\vskip 2cm
\begin{center}
{\large \bf Taming the Runaway Problem of Inflationary Landscapes} 

\vskip 1.2cm
${}^a$Lawrence J. Hall, ${}^a$Taizan Watari, and ${}^b$T. T. Yanagida

\vskip 0.4cm
${}^a${\it Department of Physics and Lawrence Berkeley National 
Laboratory,

University of California, Berkeley, CA 94720, USA} \\

${}^b${\it Department of Physics, University of Tokyo, Japan} \\

\vskip 1.5cm

\abstract{A wide variety of vacua, and their cosmological realization, 
may provide an explanation for the apparently anthropic 
choices of some parameters of particle physics and cosmology. If the 
probability on various parameters is weighted by volume, 
a flat potential for slow-roll inflation is also naturally understood, since 
the flatter the potential the larger the volume of the sub-universe.
However, such inflationary landscapes have a serious problem, 
predicting an environment that makes it exponentially hard for observers 
to exist and giving an exponentially small probability for a moderate 
universe like ours.
A general solution to this problem is proposed, and is illustrated in
the context of inflaton decay and leptogenesis, leading to 
an upper bound on the reheating temperature in our sub-universe.
In a particular scenario of chaotic inflation and non-thermal 
leptogenesis, 
predictions can be made for the size of CP violating phases, the rate 
of neutrinoless double beta decay and, in the case of
theories with gauge-mediated weak scale supersymmetry, for the
fundamental scale of supersymmetry breaking.} 

\end{center}
\end{titlepage}


\section{Introduction}

Our complicated world is based on hundreds of elements that 
have different properties.  The hundreds of elements reflect 
the existence of hundreds of stable nuclei. 
The variety of stable nuclei requires fine choices of parameters 
such as $\alpha_{\rm QCD}, \alpha_{\rm QED}, m_u, m_d$ and $m_e$. 
Even a deviation of a few \% in $\alpha_{\rm QCD}$ from the current 
value could destabilize or stabilize such nuclei as ${}^2 {\rm H}$, 
${}^2 {\rm He}$ and di-neutrons, completely changing the thermonuclear 
processes, chemical abundances of the universe and lifetime of stars
\cite{Barrow}. Such fine tuning cannot be explained by the 
naturalness principle.
Rather, it is easily explained by a combination of the many-universe 
idea (a cosmological diversity in the choice of 
theories and their parameters) with anthropic selection 
\cite{manyA,manyB}. 
It is only in the sub-universes with fine-tuned parameters 
that observers made of many species of atoms exist.
This combination also provides a simple solution to 
both the why-small and why-now problems of the cosmological constant 
\cite{CC1,CC2}.

Once we accept that the combination of many-universes with anthropic 
selection plays a role in determining the 
values of observed parameters, then we need to change 
our way of thinking. The combination not only confirms that  
the observed values of parameters are consistent with the 
existence of observers, but, because the notion of naturalness 
is replaced by probability,
it changes to some extent the questions that should be asked. 
The fine tuning of parameters and initial conditions of inflation 
is one prominent example: if the probability distribution is 
weighted by the volume of the sub-universes, such a fine tuning is 
rather probable, as long as theory space allows the inflaton mass 
to scan \cite{Vilenkin95}, or provided appropriate initial conditions 
are realized somewhere in the universe \cite{LindeChaotic}. 
We call such a space of theories containing lots of inflationary 
regions an inflationary landscape.

The problem of inflationary landscapes \cite{FHW,GV} is that we are 
no longer able to choose the parameters of the inflation model 
by hand to fit the observed data; rather these parameters 
(and consequently the normalization of the density perturbations)  
are predicted by a combination of statistics from the landscape, 
cosmological dynamics and anthropic conditions \cite{manyB}. 
In particular, it is generically true that as the  
inflaton mass becomes smaller, so the inflated volume becomes 
exponentially larger. 
This exponential behaviour of the volume-weighted 
probability distribution implies not only 
that a small inflaton mass is very probable \cite{Vilenkin95}, 
but also that the inflaton mass will become as small as
possible until it is prevented by some anthropic reason. 
Thus, a naive prediction of inflationary landscapes would be 
that the inflaton mass is so small that the environment 
  is {\it exponentially} hard for observers to exist \cite{FHW} 
because, for example, the density perturbations are too small or too large 
(this was called ``$\sigma$ problem'' in \cite{FHW}), 
the baryon asymmetry is too small or there is not enough 
matter. Whatever the reason, this picture apparently does not lead to our 
universe.\footnote{After understanding astrophysics and biology
better, however, the naive prediction may turn out to be the solution 
to Fermi's paradox ``Where is everybody?'' \cite{everybody}.}
Since the slow-roll volume factor selects vacua with smaller and smaller
inflaton mass, we call this the inflation runaway problem.

The $\sigma$ problem may be solved \cite{FHW,GV} if the density 
perturbations arise from fluctuations of light fields other than 
the inflaton \cite{curvaton,modulated}.
Even then, the inflaton mass has to be predicted 
at a moderate value, or otherwise, there could be\footnote{When the 
inflaton mass does not affect late time cosmology except for the 
number of e-foldings, as in the hybrid inflation model in \cite{FHW},
such a problem does not exist.} the same problem as above, 
for instance, 
in the baryon asymmetry or dark matter abundance.
Thus, this is quite a generic problem of inflationary landscapes.  
Note, however, that the probability distribution on observable 
parameters depends sensitively on the boundary values of unobservable 
parameters and amplitudes being scanned. Thus, the distribution is 
not exponentially sensitive to observables \cite{Linde-curvaton} 
in some landscapes where the observable--unobservable correlation 
is sufficiently week \cite{FHW}.

In this article, we present an idea of how to stop the 
runaway behaviour of the inflaton mass, along with its concrete 
implementation. 
In section \ref{ssec:rev} we introduce the runaway problem, and 
in section \ref{ssec:idea} we point out a caveat that allows
a solution. It is crucial to distinguish 
anthropically relevant parameters from fundamental parameters, and 
further 
\begin{enumerate}
 \item [(a)] 
       to have a sharp function, or threshold behaviour, of the former 
       in terms of the latter, and
 \item [(b)] 
       to have an anthropic reason that exponentially disfavours some range of 
       relevant parameters.
\end{enumerate}
The runaway problem can be overcome when both (a) and (b) are 
satisfied. In section \ref{ssec:inflatondecay}, we see that 
(a) can be realized as the reheating temperature $T_R$ depends upon 
the inflaton mass. If the inflaton decays dominantly to two heavy 
particles, $T_R$ drops rapidly as the inflaton mass becomes close 
to twice of the mass of the heavy particle. The inflaton mass is 
predicted to be close to twice the mass of the heavy particle. 
In this article we assume that only the inflaton mass and 
the cosmological constant are scanned in the landscape, not 
the mass of the heavy particle.\footnote{We do so as a first attempt 
to address the runaway problem in the simplest and easiest situation. 
There may be room to lift some of these strong assumptions, but 
this is beyond the scope of this article.}
Section \ref{sec:exp} explains how $T_R$ can be a relevant parameter.
If the baryon asymmetry is generated by thermal leptogenesis,
a very low reheating temperature leads to a low baryon asymmetry and, if the baryon 
asymmetry is too low, galaxies cannot form before protons decay. 
Thus, too low a $T_R$ is exponentially disfavoured.
Therefore, both (a) and (b) can be realized, and the runaway problem 
is solved. 
In section \ref{sec:neutrino} we study the possibility that the heavy 
particles produced by inflaton decay are identified with 
one of the right-handed neutrinos, hoping for constraints and predictions
in the neutrino sector.
The last section is devoted to conclusions and 
discussion, together with a summary of our predictions. 
The appendix investigates one of the uncertainties in 
the analysis in sections \ref{sec:exp} and \ref{sec:neutrino}, 
the effect of evaporation of particles from proto-galaxies. 

\section{The Runaway Problem and How to Solve It}
\subsection{The Inflation Runaway Problem}
\label{ssec:rev}

The probability distribution $d {\cal P}(\xi)/d\xi$ describes the 
fraction of observers in the overall universe who are in sub-universes 
where the fundamental parameters, $\xi$, are in the range 
$\xi$ to $\xi + d\xi$ \cite{Vilenkin95}. 
\begin{equation}
 d {\cal P}(\xi) = d \xi \; I(\xi) {\cal V}(\xi) {\cal A}(\alpha(\xi)).
\label{eq:distr}
\end{equation}
The initial volume distribution $I(\xi)$ includes the density of 
vacuum states and the probability distribution for the initial
amplitudes of the inflaton field. The density of states can be, 
in principle, calculated from a given landscape, or from a top-down 
theory \cite{DOS}. 
The volume factor ${\cal V}(\xi)$ \cite{Linde-eternal,volume} 
accounts for the expansion of each sub-universe during and 
after slow-roll inflation 
\begin{equation}
 {\cal V}(\xi) \propto e^{3 N_e(\xi)}.
\label{eq:vol}
\end{equation} 
With these definitions,
any volume expansion of the universe prior to the slow-roll stage 
is incorporated in the initial factor $I(\xi)$.
For instance, if a landscape supports eternal inflation, its volume 
expansion is to be incorporated in the initial factor $I(\xi)$.
The anthropic factor ${\cal A}(\alpha)$ depends only on parameters 
$\alpha$ of the low energy effective theory and of late time cosmology.
We call these the relevant parameters, some of which 
are relevant for providing an environment suitable for observers.  
Ultimately they depend on the fundamental parameters, $\alpha(\xi)$. 
The anthropic factor vanishes for many values of the relevant 
parameters, for example ${\cal A}(m_e = 5 \; \MEV) = 0$. 
However, in this paper the only aspect of the anthropic factor 
that we consider is the formation of galaxies having an appropriate 
number of baryons, since we pay attention only to the scanning of 
inflation parameters. 

It is rather likely that a landscape supports some eternally inflating 
regions \cite{Vilenkin-eternal,Linde-eternal,Susskind-eternal}; 
our standard-model vacuum with a minute cosmological constant 
may result from one of them. An eternally inflating vacuum can tunnel or jump 
into regions of the landscape that support slow-roll inflation through 
bubble nucleation or through quantum fluctuation of inflaton fields, 
eventually ending up in a radiation dominated standard-model universe 
after reheating ({\it c.f.} \cite{Susskind05}). 
The volumes of eternally inflating vacua keep expanding forever, and 
hence the volume created through nucleation or fluctuation is infinite 
as well. There is an ongoing effort to regularize these infinities 
so that the probability distribution (\ref{eq:distr}) is well-defined 
\cite{volume,regularize,GV,GV2}. 
In this article, we assume that such a regulation is possible.
We further assume\footnote{Another assumption that we make in this
article is that the volume distribution $I(\xi){\cal V}(\xi)d\xi$ 
can be treated as a continuous distribution over field theory 
parameters $\xi$. This assumption may be translated into a 
sufficiently dense landscape of vacua, or bubble tunneling rates 
without large hierarchy among them; for more information, 
see \cite{FHW}.} that, when the probability distribution is 
made well-defined, the classical volume expansion factor 
(\ref{eq:vol}) is not exactly cancelled by $I(\xi)$. Such a
cancellation appears unlikely, and one of our motivations to 
consider the scanning of some inflation  parameters is to seek an
origin for a flat inflaton potential in the volume-weighted 
probability distribution.

Since the number of e-foldings of slow-roll inflation  $N_e \simgt 60$ 
for our sub-universe, $ {\cal V}(\xi)$ ensures that the probability 
distribution involves a powerful exponential dependence 
on the parameters of the inflation potential \cite{manyB}. 
After integrating over all fundamental parameters and initial amplitudes 
of various fields that we cannot observe directly, $\xi_\parallel$, 
one obtains the probability distribution of relevant parameters 
(fraction of observers who see them) 
\begin{equation}
d {\cal P}(\alpha) = d \alpha \left[ 
   \int d\xi_{\parallel} 
         \left( d \alpha / d \xi_\perp \right)^{-1} 
         I(\xi) {\cal V}(\xi) 
                              \right] {\cal A}(\alpha).
\label{eq:probalpha}
\end{equation}
Within the anthropic window, where the anthropic factor is changing 
only mildly, the overall probability distribution is governed by 
the factor in the square bracket.
Unless there is an accidental cancellation between $I(\xi)$ and 
${\cal V}(\xi)$, or unless the region of integration determined by 
a landscape is finely designed,\footnote{A landscape approximated by 
an ensemble of chaotic inflation potentials, with a mass-independent cut off on 
the maximum field value, is such a possibility \cite{FHW}. 
Reference \cite{Linde-curvaton} presents an explicit realization.
} the distribution of observables is generically 
exponentially sensitive to inflation-related quantities such as the
density perturbations and the reheating temperature 
\cite{FHW}.
An immediate consequence is that an exponentially small fraction of 
observers in the universe would see values of these parameters 
in the middle of the anthropic window \cite{FHW,GV}. 
Since we do not accept this conclusion that our sub-universe is 
so extraordinarily special, there must be something wrong with these 
assumptions or arguments. 

In fact, the exponential behaviour from the volume expansion factor 
${\cal V}(\xi)$ is so steep that the first question to be answered 
is whether the overall probability distribution can be normalized. 
If ${\cal I}(\xi)$ and ${\cal A}(\xi)$ are power-law 
functions, the probability distribution grows forever and is
not well-defined. 
At a minimum, the exponential behaviour must be cut off so that 
observable parameters have average values that are well-defined.   
This may be accomplished in three different ways. 

If only a limited number of states are available in a landscape 
after imposing rigid anthropic conditions,\footnote{It should be 
noted that an anthropic factor from cosmological origins tends not 
to provide a rigid cut-off in the range of parameters \cite{MSW,FHW}. 
Any conditions that involve density perturbations fall into 
this category. This is because Lagrangian parameters 
(or sub-universes) determine only the standard deviation $\sigma$ 
of density fluctuations $Q$, whereas the conditions are imposed on the 
real fluctuations $Q$. The anthropic factor ${\cal A}(\sigma)$ can be 
calculated from $\widetilde{\cal A}(Q)$ by assuming a Gaussian 
distribution: 
\begin{equation}
 {\cal A}(\sigma) \approx 
   \int dQ e^{-(Q/\sigma)^2} \widetilde{\cal A}(Q). \nonumber
\end{equation} 
Thus, at most the anthropic factor can provide an exponential 
suppression.} the probability distribution is always normalizable.
On the other hand, the most probable vacuum among them does not 
necessarily satisfy non-rigid anthropic conditions \cite{FHW}.

Another possibility is that the initial volume factor provides another 
exponential distribution, so that the combined volume distribution 
$I(\xi) {\cal V}(\xi)d\xi$ has two exponentials counteracting each 
other. Various properties of the most probable vacuum will be 
determined by how the two exponential distributions balance. 
Without studying landscapes 
of the fundamental theory (including eternally inflation regions),
it may be hard to understand  
why the probable vacuum, supposed to describe our universe, happens 
to fall inside the anthropic window. 

\begin{figure}
\begin{center}
\begin{tabular}{ccc}
\includegraphics[width=.3\linewidth]{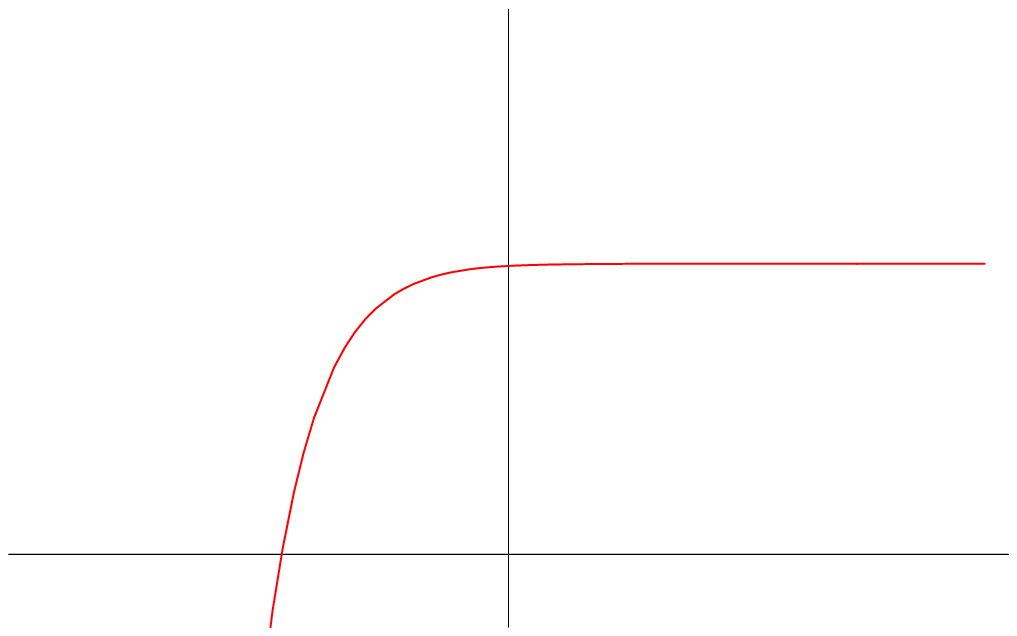} &
\includegraphics[width=.3\linewidth]{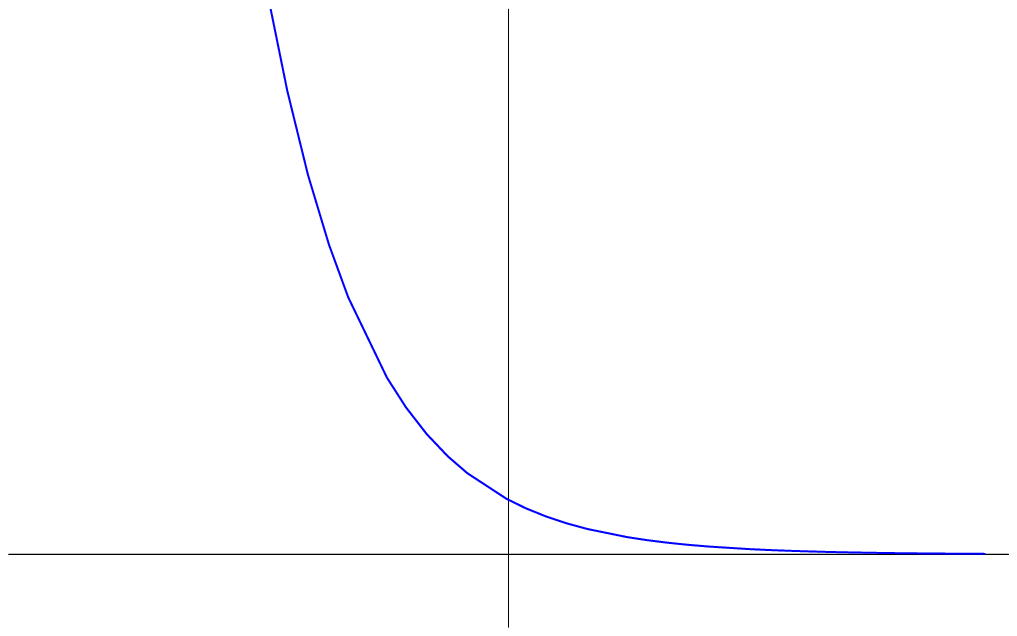} &
\includegraphics[width=.3\linewidth]{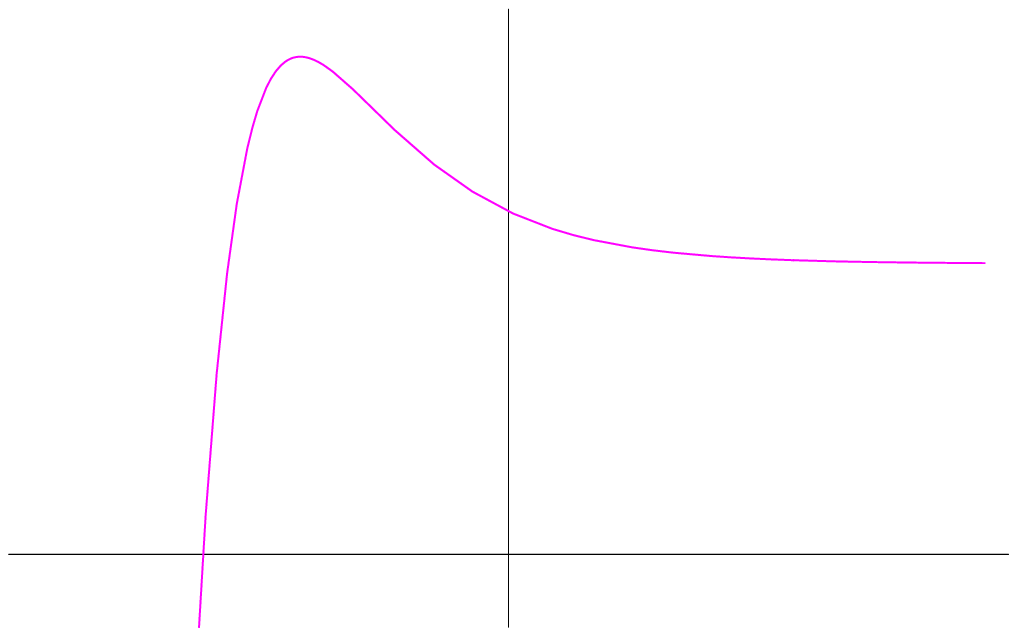} 
\end{tabular}
\begin{picture}(0,0)
\LongArrow(-340,32)(-390,32)
\Text(-380,27)[tl]{anthropic window}
\LongArrow(-25,32)(-75,32)
\Text(-65,27)[tl]{window}
\Text(-390,50)[tl]{${\cal A}$}
\Text(-235,50)[tl]{${\cal V}$}
\Text(-75,50)[tl]{${\cal AV}$}
\Text(-330,-25)[rb]{$\xi$}
\Text(-170,-25)[cb]{$\xi$}
\Text(-20,-25)[cb]{$\xi$}
\end{picture}
\caption{\label{fig:runaway}Schematic picture of the inflation runaway 
problem. From left to right: the anthropic factor ${\cal A}$, 
the volume factor ${\cal V}$, and the combined probability 
distribution ${\cal V}{\cal A}$, all on a logarithmic scale. 
Normalization of each factor is arbitrary, and hence there is no 
importance in the absolute height in the figure. 
The region to the right of the vertical axis is the anthropic window, 
and the peak of the probability distribution lies outside the window.} 
\end{center}
\end{figure}

The other possibility is that the probability is rendered well-defined 
by an anthropic factor that provides an exponential suppression. 
Instead of relying on the initial distribution, about which we 
know very little, a special form for the anthropic factor is required,
offering the hope of observable consequences. We find this case rather 
attractive and pursue it in this article. 
The problem of this possibility, however, is that it is only outside 
the anthropic window where the anthropic factor begins to decrease 
and counter the exponential growth in the inflation volume. 
If we assume that both the volume factor 
$I(\xi){\cal V}(\xi)\approx e^{3N_e(\xi_\perp)}$ and the anthropic 
factor ${\cal A}(\alpha(\xi)) \approx e^{-F(\alpha(\xi))}$ are 
governed by some power functions $N_e(\xi)$ and $F(\alpha(\xi))$, 
then the peak of the probability distribution lies roughly where  
\begin{equation}
 \frac{\partial N_e(\xi_\perp)}{\partial \ln \xi_\perp} - 
 \frac{\partial F(\alpha(\xi_\perp))}{\partial \ln \xi_\perp} \approx 
 0, \quad {\rm and~hence} \quad 
 N_e(\xi_\perp) \approx F(\alpha(\xi_\perp)).
\label{eq:trouble}
\end{equation}
This implies that the majority of observers in the universe 
live in sub-universes having an environment exponentially hard 
for observers to exist: ${\cal A}\approx e^{-F(\alpha(\xi))}\simlt 
e^{-{\cal O}(100)} \ll 1$. 
The powerful exponential factor ${\cal V}(\xi)$ pushes the expectation 
value of observables outside the anthropic window 
(see Fig.~\ref{fig:runaway}) \cite{FHW}.\footnote{
This problem persists even if density perturbations are generated by 
the curvaton \cite{curvaton} or modulated reheating \cite{modulated} 
mechanism, so that their normalization is predicted to lie within 
the window, i.e., the $\sigma$ problem is cured as in \cite{FHW,GV}. 
The density perturbations are not the only observables that have 
to be fixed within moderate values. For example, a sub-universe with a
reheating temperature sufficiently low to give an exponentially 
suppressed anthropic factor is far from the one we observe.} 
We call this the inflation runaway problem.

\subsection{An Idea to Solve the Problem}
\label{ssec:idea}

There is a caveat in the above argument that we exploit 
in this article.
The volume expansion factor is a function of the fundamental
parameters of inflation models 
$\xi$, whereas the 
anthropic factor depends on the relevant parameters, $\alpha(\xi)$. 
The caveat is in the map between them: what if the map shows 
a very sharp behaviour?

\begin{figure}[t]
\begin{center}
\begin{tabular}{cc}
 \includegraphics[width=0.45\linewidth]{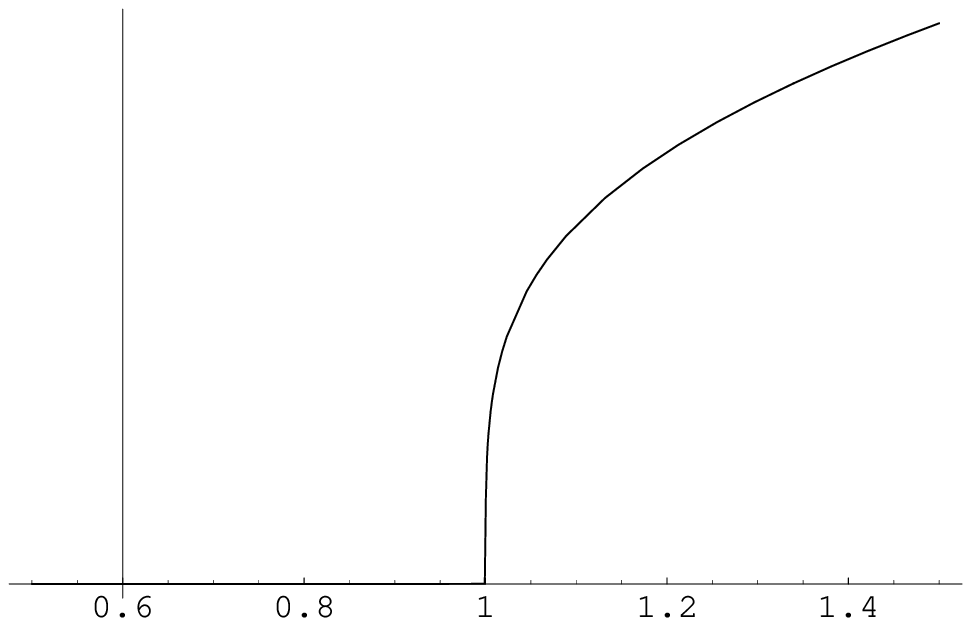} & 
 \includegraphics[width=0.45\linewidth]{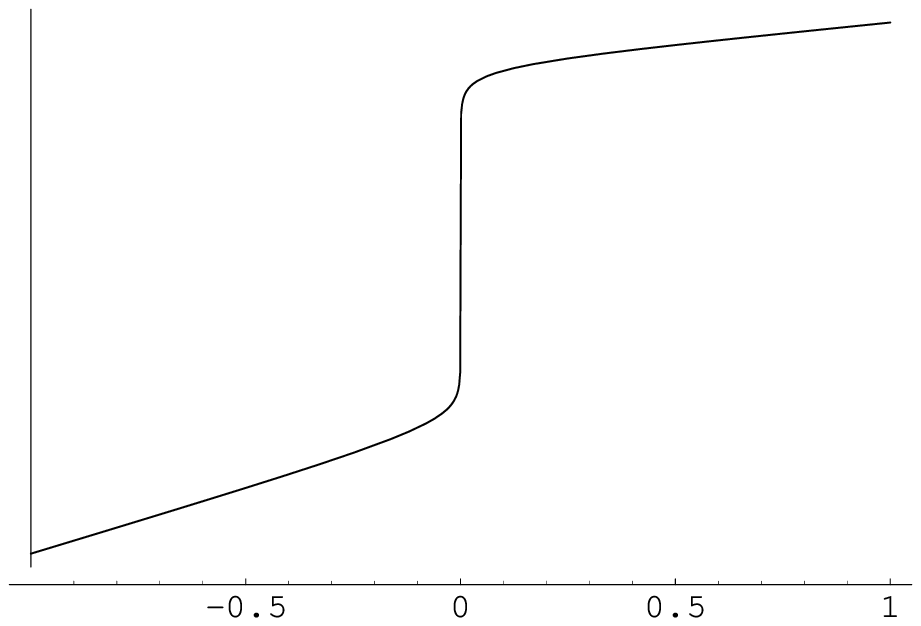} \\
\end{tabular}
\begin{picture}(0,0)
\put(-375,-72){$\xi=(m_\phi/2m_X)$}
\put(-445,80){$\alpha=T_R$}
\put(-140,-72){$\log_{10}(m_\phi/2m_X)$}
\put(-228,80){$\log_{10}(T_R)$}  
\end{picture}
\caption{\label{fig:decay} A relevant parameter $\alpha(\xi)$ with 
a sharp dependence on a fundamental parameter $\xi$. 
This is interpreted as the dependence of the 
reheating temperature on the inflaton mass, $T_R(m_\phi)$
in section \ref{ssec:inflatondecay}, and this figure 
corresponds to $\lambda' \approx 10^{-3}$.}
\end{center}
\end{figure}

Let us assume that $\alpha(\xi)$ has a region where it is rapidly 
varying, i.e., $|\alpha'| \equiv |d\alpha / d\xi_\perp|$ is large, 
as shown in Fig.~\ref{fig:decay}. Then the factor 
$(d \alpha/d\xi_\perp)^{-1}$ in (\ref{eq:probalpha}) becomes small as 
parameters enter this region, as illustrated by a green curve 
in Fig.~\ref{fig:solution},
and the volume factor ${\cal V}(\xi)$ becomes a mild 
function of $\alpha$ (blue curve 
in Fig.~\ref{fig:solution}). Although ${\cal V}$ is exponentially 
sensitive to $\xi$, the dependence on $\alpha$ in this region is 
much weaker.
Thus, the factor in the square brackets in (\ref{eq:probalpha}) can form 
a mild peak, as shown in Fig.~\ref{fig:solution}. 
If this particular region of $\alpha$ has an overlap with the 
anthropic window, where ${\cal A}(\alpha)$ depends only mildly on 
$\alpha$, the combined probability distribution may have a mild 
peak inside the anthropic window, as shown in the right of 
Fig.~\ref{fig:solution}. 
This explains how inflationary landscapes can predict average 
values of observables inside the anthropic window.

In this case, the difficulty in (\ref{eq:trouble}) is avoided in 
the following way. The derivative of the distribution 
function in (\ref{eq:probalpha}) has to be zero somewhere 
in the anthropic window, forming a peak in the distribution. There, 
\begin{equation}
 - \frac{\partial \ln \alpha'}{\partial \ln \alpha} 
 + \frac{\partial \ln \xi_\perp}{\partial \ln \alpha}
   \frac{\partial \ln N_e}{\partial \ln \xi_\perp}\bar{N}_e
 + \frac{\partial \ln {\cal A}}{\partial \ln \alpha} = 0,
\label{eq:trouble?}
\end{equation}
where $\bar{N_e}$ is the number of e-foldings of a sub-universe 
of maximum probability.  If $\alpha'$ is a power function of $\alpha$,
the first term is of order unity. If $N_e$ and $\ln {\cal A} \sim - F$
are simple power functions of $\xi_\perp$ and $\alpha$, respectively, 
$\partial \ln N_e / \partial \ln \xi_\perp$ is also of order unity, 
and the last term is of order $-F(\alpha)$. Since we want our own 
sub-universe to be a highly probable one, $\bar{N}_e$ is approximately 
the number of e-foldings in our sub-universe, which is much larger 
than one. On the other hand, we would like our vacuum to be predicted 
inside the anthropic window, and hence $F(\alpha)$ should be of order unity. 
Thus, the second term must also be of order unity at the peak of the 
distribution, in spite of the large value of $N_e$. This is possible  
due to the sharp distribution, as long as 
\begin{equation}
\frac{\partial \ln \alpha}{\partial \ln \xi_\perp} \approx 
\bar{N}_e 
\label{eq:sensitivity}
\end{equation}
is satisfied.
The more the sub-universe inflates, the sharper is the required 
behaviour of $\alpha(\xi)$.

\subsection{A Mass Threshold in Inflaton Decays}
\label{ssec:inflatondecay}

\begin{figure}
\begin{center}
\begin{tabular}{cc}
\includegraphics[width=0.45\linewidth]{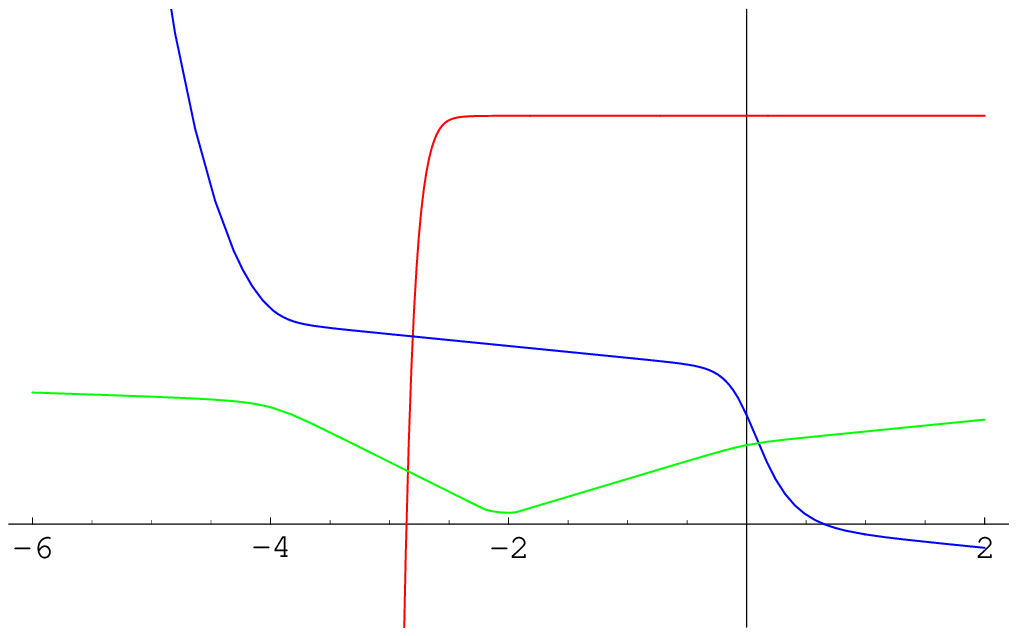} &
\includegraphics[width=0.45\linewidth]{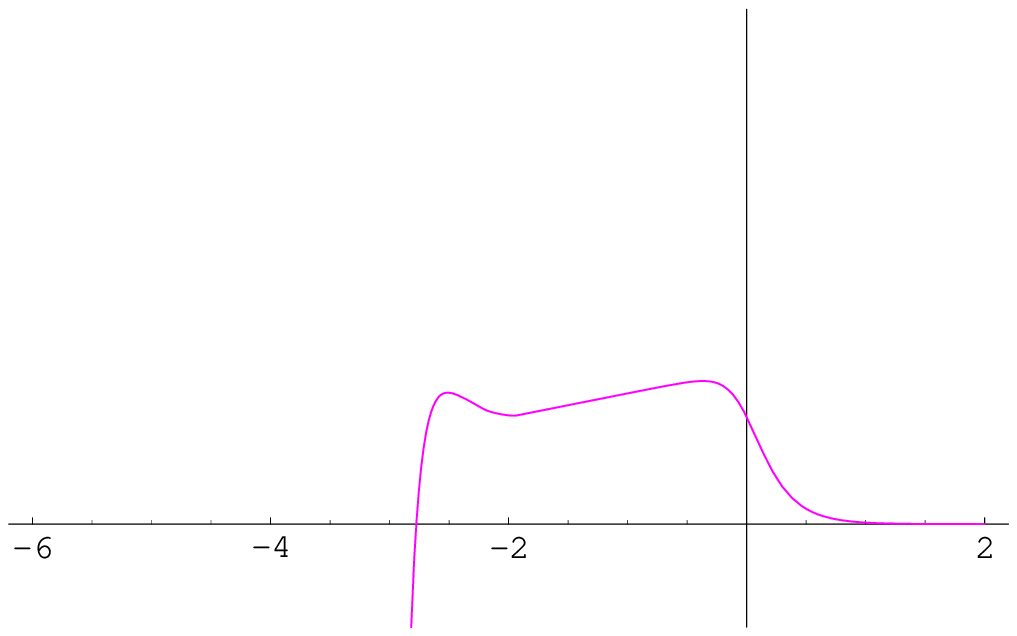}
\end{tabular}
\begin{picture}(0,0)
\put(-285,-60){$\log_{10}(T_R/T_+)$} 
\put(-60,-60){$\log_{10}(T_R/T_+)$}
\LongArrow(-350,55)(-300,55)
\LongArrow(-330,55)(-400,55)
\Text(-350,58)[cb]{threshold region}
\LongArrow(-330,40)(-360,40)
\Text(-325,40)[l]{anthropic window}
\Text(-275,52)[cb]{${\cal A}$}
\Text(-420,50)[r]{${\cal V}$}
\Text(-430,-10)[b]{$(\alpha^{'-1})$}
\LongArrow(-125,55)(-75,55)
\LongArrow(-105,55)(-175,55)
\Text(-125,58)[cb]{threshold region}
\LongArrow(-105,40)(-135,40)
\Text(-100,40)[l]{anthropic window}
\Text(-75,15)[cb]{$T_{R,0}$}
\LongArrow(-75,13)(-75,-5)
\Text(-135,-5)[cb]{$(d{\cal P}/d\alpha)$}
\end{picture}
\caption{\label{fig:solution}Left: various factors of the probability 
distribution, namely, ${\cal A}, {\cal V}$ and $(d \alpha/d\xi)^{-1}$, 
on a logarithmic scale. The normalization of each factor is arbitrary, and 
hence there is no importance to the absolute height.
Right: combined probability distribution $d{\cal P}/d\alpha$. 
Note that the peak of the distribution is inside the anthropic window.
[In this figure, the plateau of the volume factor in the threshold 
region corresponds to $\bar{N}_e \sim 10$. This is why the peak 
$T_{R,0}$ is close to the upper end of the threshold region.]
}
\end{center}
\end{figure}

After inflation the universe must reheat, so that the inflaton must decay. 
We assume that the inflaton mass scans ($\xi = m_\phi$) and that
${\cal V}(m_\phi)$ (or the combination $I(\xi){\cal V}(\xi)$) increases 
as $m_\phi$ decreases. As the inflaton mass $m_\phi$ is scanned, 
mass thresholds in the inflaton decay rate automatically provide 
a sharp behaviour in the reheating temperature $T_R$: 
thus we take $\alpha(\xi)$ to be a single relevant parameter, 
the reheating temperature $T_R(m_\phi)$.
Suppose the inflaton $\phi$ decays dominantly to a pair of heavy fermions, 
$\phi \rightarrow \psi_X + \psi_X$, through the operator  
\begin{equation}
{\cal L} = \frac{1}{2} \lambda \phi \psi_X \psi_X + {\rm h.c.},
\label{eq:decay1}
\end{equation}
and that the universe is reheated by $\psi_X$ decaying to a fermion 
$\psi_Y$ and a scalar $Y$ through 
\begin{equation}
 {\cal L} = \lambda' \psi_X \psi_Y Y + {\rm h.c.}.
\label{eq:decay2}
\end{equation} 
For  $m_{\phi} > 2 m_X$, the 2-body decay rate of the inflaton is 
\begin{equation}
 \Gamma(\phi \rightarrow \psi_X + \psi_X) = 
   \frac{1}{16\pi}\lambda^2 m_{\phi} 
   \sqrt{\beta},
\end{equation}
where $\sqrt{\beta}$ describes the mass threshold, and 
$\beta \equiv 1-(2m_X/m_\phi)^2$ decreases sharply as the inflaton 
mass $m_\phi$ approaches the threshold $2m_X$ from above. 
The 2-body decay channel closes when $m_{\phi} < 2m_X$, 
forcing the inflaton to decay through a 3-body channel 
\begin{equation}
 \phi \rightarrow \psi_X +\psi_X^* \rightarrow \psi_X + \psi_Y + Y,
\end{equation}
where $\psi_X^*$ is virtual, with a decay rate 
\begin{equation}
 \Gamma(\phi \rightarrow \psi_X + \psi_Y + Y) \approx 
   \frac{1}{16\pi}\left(\frac{2\lambda^{'2}}{16\pi^2}\right)\lambda^2 
   \frac{m_{\phi}^3}{m_X^2}
\label{eq:3body-rate}
\end{equation}
for $m_X < m_\phi < 2m_X$. The decay rate of the inflaton suddenly 
decreases by roughly $(\lambda'/4\pi)^2$ as the inflaton crosses 
the threshold at $m_{\phi} \simeq 2m_X$. Since the reheating temperature 
$T_R$ is given by $\sqrt{\Gamma M_{\rm pl}}$, it drops rapidly 
\begin{eqnarray}
 {\rm from}& T_+ \equiv & {\rm min}(\lambda,\lambda') 
                           \sqrt{m_{\phi} M_{\rm pl}/16\pi}, 
\label{eq:Tp} \\
 {\rm to} & T_- \equiv & \frac{\lambda \lambda'}{4\pi} 
                           \sqrt{m_{\phi} M_{\rm pl}/16\pi},
\label{eq:Tm}
\end{eqnarray}
as $m_{\phi}$ passes through the threshold at $2m_X$, 
as shown in Fig.~\ref{fig:decay}. 
This illustrates the origin of the sharp behaviour in $\alpha(\xi)$: 
the late-time cosmology parameter $\alpha$ is $T_R$  
and the high-energy parameter $\xi$ is the inflaton mass, $m_\phi$.
Of course the idea in section \ref{ssec:idea} will work only when 
the anthropic factor depends exponentially onn $T_R$, so that we 
will be led to consider theories where the baryon asymmetry and/or 
the  dark matter relic density depend on $T_R$.

Let us examine more explicitly how the averaged value of 
low-energy observables are determined in the example of 
$\alpha(\xi) = T_R(m_{\phi})$ above. The threshold behaviour 
of $T_R$ is approximated\footnote{The approximation of $T_R$ using 
3-body decay also exhibits sharp behaviour because 
the virtual $\psi_X$ is almost on shell as $m_{\phi}$ approaches 
the threshold from below. The definition of $\beta$ is
$2[(m_{\phi}/2m_X)-1]$ around the threshold.} by 
\begin{equation}
 T_R (m_{\phi})|_{m_{\phi} \approx 2m_X} \approx \left\{ 
\begin{array}{lc}
 T_+ \beta^{1/4} & 
 \left[ m_{\phi} >  2m_X (1+\lambda^{'2}/(16\pi)) \right], \\
 T_- |\beta|^{-1/4} & 
 \left[ m_{\phi} < 2m_X (1-\lambda^{'2}/(16\pi)) \right]. 
\end{array}
\right. 
\end{equation}
Here, we have assumed that the reheating process above the threshold 
is governed by the decay of the inflaton $\phi$, rather than the decay 
of $\psi_X$.
We focus on the part of the threshold with $m_\phi > 2 m_X$, and
substitute $\alpha(\xi) = T_R (m_\phi) = \beta^{1/4} T_+$ in
(\ref{eq:sensitivity}) to give
\begin{equation}  
T_{R,0} \approx \frac{T_+}{\bar{N}_e^{1/4}}
\label{eq:TR}
\end{equation}
for a highly probable universe, such as our own.
This is an important result. Our mechanism requires the reheat
temperature of our sub-universe to lie in the upper half\footnote{
Here, we assume that the volume factor increases as the inflaton mass 
decreases, as in Fig.~\ref{fig:solution}. There may be another peak 
in the lower half of the threshold region, as in 
Fig.~\ref{fig:solution}, but it depends on all the power-law 
components of all the factors in the probability distribution whether 
this peak exists or not, and which peak is higher. We just ignore 
this peak in this article. } of the
threshold region, so that $T_{R,0} > \sqrt{T_+ T_-} \approx 
\sqrt{{\rm max}(\lambda,\lambda')/(4\pi)}T_+$. 
However, even for a huge volume factor ${\cal V}$, 
this requirement is very mild: 
${\rm max}(\lambda,\lambda') < 4 \pi / \bar{N}_e^{1/2}$. 
In the following sections we find that a sufficient exponential 
suppression of the anthropic factor will provide a somewhat stronger 
constraint on max($\lambda,\lambda'$). 

It is instructive to see in more detail how the probability
distribution is maximized in this example of a mass threshold in
inflaton decays.
Assuming there is no significant structure in 
$I(m_{\phi}) d m_\phi$ around $m_{\phi} \approx 2m_X$, the initial 
distribution is approximated by 
\begin{equation}
 I(m_{\phi}) dm_{\phi} \propto d (m_{\phi}-2m_X) \propto 
 \left(\frac{T_R}{T_+}\right)^4 d\ln \left(T_R/T_+ \right).
\end{equation}
Meanwhile, assuming that the e-fold number $N_e(m_{\phi})$ 
does not have significant structure around $m_{\phi} \simeq 2m_X$, 
we adopt the Taylor expansion 
\begin{equation}
 N_e(m_\phi) \approx N_e(2m_X) \times 
  \left(1 - c \left(\frac{m_{\phi}-2m_X}{2m_X}\right)\right)
  \approx \bar{N}_e (1 - c \beta),
\end{equation}
with an order one coefficient $c$. Here $N_e$ is assumed to  
increase as $m_\phi$ decreases (as in the case of landscapes 
of chaotic-inflation regions \cite{Vilenkin95,FHW,GV}), 
hence the minus sign.
The volume expansion factor is now approximated by 
\begin{equation}
 {\cal V}(m_{\phi}) \approx e^{3 \bar{N}_e} 
  e^{-3c \bar{N}_e (T_R/T_+)^4}.
\end{equation}
The overall probability distribution (ignoring the contribution from 
the anthropic factor in the anthropic window) is
\begin{equation}
{\cal V}(m_{\phi})I(m_{\phi}) d m_\phi 
  \approx e^{3 \bar{N}_e} e^{-3c \bar{N}_e (T_R/T_+)^4} 
  \left(\frac{T_R}{T_+}\right)^4 d \ln (T_R/T_+)
\end{equation}
has a peak around $T_R/T_+ \approx (1/ \bar{N}_e)^{1/4}$, reproducing
(\ref{eq:TR}). 
This is how the typical $T_R$ is determined from this distribution, 
and will be the reheating temperature of our sub-universe $T_{R,0}$ if 
our vacuum is to be typical. The inflaton mass $m_\phi$ is 
predicted in this framework almost at the threshold $2m_X$ with 
a deviation of order $(T_{R,0}/T_+)^4 \approx 1/ \bar{N}_e$, and the 
reheating temperature is suppressed by $(1/ \bar{N}_e)^{1/4}$ from
the naive expectation $T_+$ because the inflaton mass is very 
close to the threshold. 

The above argument shows how the probability distribution can form 
a peak within the anthropic window. We need to further make sure that 
this is the only peak, or at least, the biggest peak in the probability 
distribution. For $m_\phi \ll m_X$, $T_R(m_\phi)$ is not a particularly 
sharp function. $T_R$ different by a factor of a few would be 
a result of $m_\phi$ different by a similar degree, which means 
the number of e-folding different by a similar degree. Increase of 
$N_e$ by a factor of 2 implies that the volume factor is multiplied by 
an extra factor of $e^{3N_e}$, since $e^{3N_e \times 2}=e^{3N_e}e^{3N_e}$.
Thus, we still need something that counters this exponential growth 
of probability distribution, and we expect the anthropic factor to do 
this job; ${\cal A}(\alpha) \approx e^{-F(\alpha)}$ with $\alpha=T_R$ 
now. 
If both $N_e(\xi)$ and $F(\alpha(\xi))$ are approximated by simple 
power functions of $\xi=m_\phi$ below the threshold, or $T_R < T_-$,
it is sufficient to make sure that 
\begin{equation}
 F(T_-) \simgt \bar{N}_e
\label{eq:atTmin}
\end{equation}
and that $F(m_\phi)$ is increasing faster than $N_e(m_{\phi})$ as 
$m_\phi$ 
decreases, in order to guarantee that there is no other peak outside 
this threshold region.

In the following sections, we discuss what kind of cosmological 
scenarios can provide such an exponential anthropic factor 
${\cal A}(T_R) \approx e^{-F(T_R)}$.

Independently of particular realization of ${\cal A}(T_R)$, however, 
one can see the following. It follows from the requirement 
$F(T_-) \simgt \bar{N}_e$ that $T_-/T_+$ is sufficiently small. 
This means that there is an upper limit on 
\begin{equation}
 \frac{T_-}{T_+} = \frac{{\rm max}(\lambda,\lambda')}{4\pi},
\end{equation}
and hence on 
\begin{equation}
 \frac{T_+}{\sqrt{m_\phi M_{\rm pl}/(16\pi)}} = 
   {\rm min}(\lambda,\lambda') \leq {\rm max}(\lambda,\lambda').
\end{equation}
Thus, the reheating temperature of our universe $T_{R,0}$ given by 
(\ref{eq:TR}) is also bounded from above, once $\bar{N}_e$ and 
$m_\phi$ are known.
Although the upper bound itself depends on such issues as particular 
cosmological scenarios that determine $F(T_R)$, $\bar{N}_e$, and 
$m_{\phi}$, yet it is generic that we have an upper bound on 
$T_{R,0}$ for individual cosmological scenario. 
We see a particularly important consequence in the scenario 
in section \ref{sec:neutrino}.

In this article, we consider the scanning of a single parameter
of the inflaton potential that causes both $m_\phi$ and $N_e$ to
vary. 
It is beyond the scope of this article to explore the possibility 
of solving the runaway problem for more complicated inflationary 
landscapes with more than one scanning inflationary parameters.
We just note here that the scanning of multiple inflationary parameters 
generically leads to exponential behaviour in multiple directions 
on the space of observable parameters, requiring multiple
anthropic reasons for exponential suppression in the number of observers. 

Our analysis in this article is based on an assumption that 
only two parameters are scanned in a landscape: the cosmological 
constant and the mass of the inflaton. All other parameters 
are assumed not to be effectively scanned: some parameters may not 
be scanned in the landscape \cite{ADK,Acharya}, or some of 
the parameters are scanned in the landscape but have already been 
pinned down to the value of our sub-universe via some 
anthropic argument. 
The two parameter scanning of $(m_\phi, \Lambda^4)$ is certainly 
minimal in considering the runaway problem, and at least the 
first place to start.

\section{An Exponential Anthropic Factor}
\label{sec:exp}

The two crucial ingredients for solving the runaway problem are 
to have a sharp threshold behaviour of a relevant parameter as 
the inflaton mass changes, and to have the anthropic factor 
${\cal A}$ decreasing exponentially with this relevant parameter---not 
just as a power law. Since we have seen in the previous section that 
the reheating temperature exhibits a sharp behaviour as the 
inflaton mass passes a threshold $2m_X$, it would be sufficient 
if, in some range of parameters, ${\cal A}$ depends exponentially
on $T_R$. In the following sections, we show how this occurs in 
some cosmological scenarios.

The basic idea is the following. The baryon symmetry $Y_B \equiv 
n_B/s$ decreases in some cosmological scenarios if $T_R$ decreases; 
thermal leptogenesis \cite{BPY} is an example---there is insufficient
production of right-handed neutrinos in the thermal plasma if $T_R$ 
is less than the mass of the lightest one. Although the number of baryons in a given 
comoving volume determines the maximum number of galaxies available, 
and hence contributes to the anthropic factor, the baryon 
asymmetry itself is usually only a power-law function of $T_R$, 
and is not enough to counter the exponential growth of the volume 
factor. Some other physics consideration is necessary in order for 
$T_R$ to be exponentially relevant to the anthropic factor.

After density perturbations grow during the era of matter domination, 
baryons begin to self-gravitate to form bound states.
In order for these bound states to evolve into such highly non-linear 
structures as galaxies, the systems must release energy.
When the number density of baryons is too low, the cooling through 
bremsstrahlung is so inefficient that the time scale for the cooling 
becomes longer than the lifetime of a proton. This observation 
leads to an exponentially suppressed anthropic factor, as shown below 
in more detail.

The baryon asymmetry remains constant in thermal leptogenesis 
as long as $T_R > M_1$, where $M_1$ is the mass of the lightest 
right-handed neutrino:
\begin{equation}
 Y_{B,0} \equiv \left(\frac{n_B}{s}\right)_{T_R > M_1} 
 \simeq 0.35 \times 
  \left(\frac{\Gamma_{N_1}}{M_1^2 / M_{\rm pl}}\right) 
  \epsilon_{\rm CP},
\label{eq:DnL-standard}
\end{equation}
where $\Gamma_{N_1}/(M_1^2/M_{\rm pl})$ is $n_{N_1}/s$, assuming 
out-of-equilibrium decay and ignoring $g_*$, the effective statistic
degrees of freedom. The CP asymmetry $\epsilon_{\rm CP}$ is given by 
\begin{equation}
 \epsilon_{\rm CP} = \frac{3}{16\pi} 
  \frac{{\rm Im}((h_{1\alpha} h_{j\alpha}^*)^2 M_1/M_j)}
       {|h_{1\alpha}|^2}.
\label{eq:CP}
\end{equation}
As $T_R$ drops below $M_1$, however, the baryon asymmetry 
from thermal leptogenesis is much less than (\ref{eq:DnL-standard}), 
because leptogenesis stops before reheating completes. After the end
of the slow roll era, the inflaton field oscillates and
the temperature of the thermal plasma behaves as 
$T^4 \sim T_R^2 \rho_{\rm infl.}^{1/2}$ \cite{KT}. 
Thermal leptogenesis stops roughly when $T \sim M_1$, and 
the subsequent entropy production continues until $T \sim T_R$, 
diluting the lepton asymmetry by a factor of $(T_R/M_1)^5$ \cite{KT}.
Furthermore, the number density of right-handed neutrinos 
is less than usual because 
\begin{equation}
 \left(\frac{n_{N_1}}{s}\right)_{T=M_1} \sim \Gamma_{N_1} t_{T=M_1} 
 \sim \left(\frac{\Gamma_{N_1}}{\rho_{\rm infl}^{1/2} /M_{\rm pl}}
      \right)_{T=M_1} 
 \sim \frac{\Gamma_{N_1}}{(M_1^2 / M_{\rm pl})} 
      \left(\frac{T_R}{M_1}\right)^2.
\end{equation}
Thus, the baryon asymmetry decreases as 
\begin{equation}
 Y_B \simeq Y_{B,0} \left(\frac{T_R}{M_1}\right)^7
\label{eq:7th-power}
\end{equation}
for 100 GeV $< T_R < M_1$.
The baryon asymmetry goes down rapidly\footnote{An implicit assumption 
here is that the branching ratio of the inflaton decay to 
right-handed neutrinos is small enough. In particular, 
none of the particles $\psi_X$, $\psi_Y$ and $Y$ in the model of 
threshold behaviour in the previous section can be identified 
with the right-handed neutrinos.} as $T_R$ decreases, 
and so does the number of Milky-Way type galaxies in a given comoving 
volume. However, since this is only a power law of $T_R$, 
this is not enough to produce the desired exponential anthropic factor. 

Let us assume that the dark matter density of our universe does 
not depend on $T_R$ (when $T_R$ is in the range, say, above the 
electroweak scale). Thermal relics of stable particles that weigh 
of order 100 GeV (WIMPs) are a good enough candidate. 
Under this assumption, the epoch of matter-radiation equality 
remains almost unchanged.
On the other hand, a single Milky-Way type galaxy has a 
fixed\footnote{The total mass of a galaxy is an important parameter 
of environment for the existence of observers. At least, the 
gravitational potential of a galaxy must be deep enough so that 
a supernova explosion cannot blow up the galaxy. Furthermore, 
correlations have been found observationally between luminosity 
(and mass) and metalicity of galaxies, and between the metalicity 
and sizes and orbiting radii of planets \cite{Planets}. 
Thus, there will be a band (or at least lower bound) in the mass of 
galaxies suitable for the existence of observers, once all other 
cosmological and particle-physics parameters are fixed. 
In our universe, it will contain $10^{11}\mbox{--}10^{12} M_{\odot}$, 
and the lower bound of the band will be no less than 
$10^8\mbox{--}10^9 M_{\odot}$---the constraint from the supernova 
explosion. 
The band (or the lower bound) itself will be different from ours 
in a sub-universe different from ours. For instance, the supernova 
constraint will be $M_{\rm gal}\simgt M_{\rm gal,SN.min.}$, where 
$M_{\rm gal,SN.min.} (T_{\rm vir}/m_p) \approx 10^{-3} M_{\odot}$ 
with the virial temperature at the epoch runaway cooling starts. 
Constraints from the metalicity will require further astrophysical 
understanding. 
Because we do not know the anthropic band (lower bound) in universes 
quite different from  ours, in this article we simply use the estimate
$M_{\rm gal} \sim (10^{11}\mbox{--}10^{12}) M_{\odot}$.
These enormous astrophysical uncertainties, however, do not change 
one of our main claims that an exponential 
anthropic factor follows by requiring galaxies to form before protons 
decay, providing the anthropic band (lower bound) of galaxy masses 
does not decrease as fast as $Y_B^{3/2}$.
Our quantitative predictions for the upper bound of the reheating 
temperature are not greatly affected either. Even if there are 
uncertainties of several orders of magnitude in the anthropic bound 
on $Y_B$ in (\ref{eq:bound-1}) or (\ref{eq:bound-2}), the limit on 
$T_R$ is changed very little because the baryon asymmetry depends 
so sensitively on $T_R$ in (\ref{eq:7th-power}) or (\ref{eq:5th-power}). 
See also section \ref{sec:concl}.}
baryonic mass, of order $M_{\rm gal.} \sim 
\left(10^{11}\mbox{--}10^{12}\right) M_{\odot}$, 
and hence a galaxy has to collect baryons from much wider comoving 
volume in sub-universes having small baryon asymmetry. The galactic 
comoving scale enclosing a baryon of order $M_{\rm gal}$
enters the horizon later if the baryon asymmetry is smaller.
The horizon entry becomes as late as the epoch of matter-radiation 
equality if $Y_B$ were\footnote{
For $\Omega_{CDM}h^2 \simeq 0.113$ and $\Omega_B h^2 \simeq 0$, 
$T_{eq} \sim 0.63 \; \EV$. 
} of order $\sim (M_{\rm gal}T_{eq}^3/m_p M_{\rm pl}^3) \sim 
(10^{-15}\mbox{--}10^{-14})(T_{eq}/0.6 \; \EV)^3$. 
For $Y_B$ less than that, the galactic comoving scale would be 
proportional to $Y_B^{-1/3}$, the horizon-entry time to 
$Y_B^{-1}$, 
and the number density of the baryon asymmetry [photons] at horizon 
entry to $Y_B^3$ [to $Y_B^2$]. This number density is diluted 
by $Q^3$ as the universe continues to expand, until the density 
perturbation of the galactic comoving scale goes non-linear. 
The density in over-dense regions increases further by of order 
$100$ during the virialization process, and the evolution of 
gravitationally bound systems decouple from the expansion of the 
rest of the universe,\footnote{
The cosmological constant has to be small enough, so that the universe 
remains matter dominated until this epoch. This only leads to a 
power-law suppression in the number of states in landscapes, and we ignore 
such a power-law contribution to the probablity 
distribution.} giving proto-galaxies with 
\begin{eqnarray}
 \rho & \approx & 
    100 T_{eq}^4 (Y_B/10^{-14.5})^2 (0.6\;\EV/T_{eq})^6 Q^3, 
   \label{eq:rho}\\
 n_\gamma & \approx & 
    (T_{eq} Q)^3 (Y_B/10^{-14.5})^2 (0.6\;\EV/T_{eq})^6, 
   \label{eq:n-gamma}\\
 n_B & \approx & 1000 n_\gamma Y_B,\label{eq:n-B} 
\end{eqnarray}
at the epoch of virialization for $Y_B < (10^{-15}\mbox{--}10^{-14})
(T_{eq}/0.6 \; \EV)^3$. 
Since the galactic comoving scale enters the horizon after
matter-radiation equality for such low baryon asymmetry, the 
virial temperature of baryons is roughly given by \cite{TR}
\begin{equation}
 T_{\rm vir} \sim m_p Q, 
\label{eq:vir}
\end{equation}
which is of order 10 keV for $Q \sim 10^{-5}$.

For sub-universes with such a low baryon asymmetry, the condition for 
runaway cooling \cite{OR} of the proto-galaxies,
\begin{equation}
 \Gamma_{\rm cool} > \left(G_N \rho \right)^{1/2},
\label{eq:cooling}
\end{equation}
is not satisfied. Here, $\Gamma_{\rm cool}$ is defined in 
terms of the energy-loss rate per particle 
\begin{equation}
 d E_{\rm loss}/dt = - \Gamma_{\rm cool} T_{\rm vir}. 
\end{equation}
Proto-galaxies can release energy 
by emitting photons. The contributions to 
$\Gamma_{\rm cool}$ from the two relevant processes, namely  
bremsstrahlung $p + e \rightarrow p + e + \gamma$ and 
Compton scattering
$e^- + \gamma \rightarrow e^- + \gamma$, are given by \cite{Formula} 
\begin{eqnarray}
 \Gamma_{\rm Brems} & \approx & 9
    \frac{\alpha^3}{m_e^2} \frac{n_B}{\sqrt{T_{\rm vir}/m_e}}, 
                \label{eq:brems}\\
 \Gamma_{\rm Comp} & \approx & \frac{2^7\pi^3}{810}
       \frac{\alpha^2}{m_e^2} \frac{T_\gamma^4}{m_e}, 
\label{eq:Comp}
\end{eqnarray}
where $T_\gamma \sim n_\gamma^{1/3}$ is the typical photon energy.
To see that (\ref{eq:cooling}) is not satisfied,  
first note that Compton scattering is more efficient 
than bremsstrahlung right after virialization;
second, the rate (\ref{eq:Comp}) does not satisfy (\ref{eq:cooling}).

Although \cite{TR} used the condition (\ref{eq:cooling}) as an 
anthropic criterion, the cooling condition (\ref{eq:cooling}) 
not being satisfied only implies\footnote{The size of the cosmological 
constant determines the largest structures with order one density 
perturbations, but it is more appropriate to think that the upper bound
on the cosmological constant is set by requiring that galaxies in the 
anthropic mass band be formed. Since the landscape picture 
prefers the largest anthropically allowed cosmological constant, 
it is possible that structures containing baryons of order 
$M_{\rm gal}$ (or the lower anthropic mass bound for galaxies) are the 
largest. If so, the destruction of proto-galaxies through mutual 
gravitational interactions before cooling starts is no longer 
an anthropic problem \cite{OR}. See, however, also footnote 
\ref{fn:cloud} and section \ref{sec:concl}.} 
that runaway cooling does not happen right after virialization. 
Instead, quasi-static cooling occurs on a time scale of order 
$\Gamma_{\rm cool}^{-1}$,
allowing gravitational contraction. During quasi-static 
cooling, the gravitationally bound system maintains virial 
equilibrium, and the Jeans mass remains equal to the total mass 
of the bound system \cite{OR}. Hierarchical cooling 
\cite{hierarchy} does not occur. Thus, sub-structures of galaxies, 
such as clumps of molecular clouds or hydrogen-burning stars,  
are not formed during quasi-static cooling.
As the quasi-static cooling proceeds, the system becomes more 
dense and hotter,\footnote{At an earlier stage of gravitational 
contraction, the source of gravitational potential energy is dominated 
by dark matter, and thus the temperature of baryons may not increase 
as in the standard $T_B \propto \rho_B^{1/3}$ relation.  
But after a baryonic core with $\rho_B \sim \rho_{\rm CDM}$ 
is formed, as a result of gravitational contraction of baryons, the 
temperature of the baryons (including electrons) begins to increase.} 
and eventually the cooling condition (\ref{eq:cooling})
is satisfied, so that hierarchical cooling starts, and 
smaller structures begin to grow.\footnote{It is the density 
at this epoch that should be used in evaluating the anthropic 
conditions in \cite{too-dense}, ensuring that the halo not be too 
dense. See section \ref{sec:concl}.} 
Thus, we do not find it really safe to claim that the anthropic 
factor is exponentially suppressed 
whenever the cooling condition (\ref{eq:cooling}) is not satisfied.

The cooling time scale cannot be arbitrary 
long;\footnote{\label{fn:cloud} 
Main-sequence stars and planetary systems around them 
are considered to have formed from cold clouds in 
galactic discs in our universe. Not all the particles are heated up 
to the virial temperature at the beginning of virialization, as 
argued in \cite{pancake}. But, if it takes too long time for  
proto-galaxies to cool, would-be discs may be heated 
and evaporate before runaway cooling starts. This anthropic 
condition is not taken into acount in section \ref{sec:exp} and 
\ref{sec:neutrino}. See discussion in section \ref{sec:concl}.}
galaxies must form before protons decay, which we assume
happens with a lifetime of order $\tau_p \approx 10^{36}$ years:
\begin{equation}
 \Gamma_{\rm cool}^{-1} < \tau_p\approx 10^{36} {\rm yrs}.
\label{eq:proton}
\end{equation}
Since relativistic particles, namely photons, cannot be trapped 
in a gravitational potential well without sufficient scattering 
interactions, the number density of photons continues to decrease 
because of the expansion of the entire universe.
On the other hand, the number density of protons remains constant 
in the virialized systems. Thus, the cooling time scale is set 
by the bremsstrahlung process.\footnote{Here, we assume that all the 
particles in proto-galaxies remain bounded gravitationally forever. 
However, this is not really correct; some particles evaporate from 
proto-galaxies before runaway cooling starts. See 
discussion in the appendix for the effects of evaporation.} 
Using (\ref{eq:n-gamma})--(\ref{eq:vir}) and (\ref{eq:brems}) in 
the anthropic condition (\ref{eq:proton}), we see that 
\begin{equation}
 \left(\frac{10^{-20}}{Y_B}\right)^3 
 \left(\frac{2\times 10^{-5}}{Q}\right)^{2.5} \simlt 1.
\label{eq:bound-1}
\end{equation}

If $Y_B \simlt 10^{-18}$, however, the $\bar{p}p$ pair annihilation
in the early universe is incomplete,\footnote{TW thanks 
G.~Perez for bringing this issue to our attention.} so that
$\bar{p}$ survive with an abundance 
of order $n_{\bar{p}}/s \sim 10^{-18}$, which, by charge conservation,
is also the $e^+$ abundance. 
The bremsstrahlung cooling rate should be evaluated by using 
$n_{p,\bar{p}}$ in (\ref{eq:brems}), instead of $n_B$.
But, since the cross section of $e^+ e^-$ pair annihilation is  
larger than that of $(e,p)$ bremsstrahlung, pair annihilation 
starts before cooling. As the number densities of electrons and 
positrons decrease, the $(e,p)$ bremsstrahlung cooling rate also 
decreases; the total energy carried by electrons (and 
positrons) is only $n_{e,\bar{e}}/n_{p,\bar{p}}$.
In the end, only electrons are left with the abundance 
$n_e = n_p-n_{\bar{p}}$, and $\Gamma_{\rm Brems}$ reduces to the one 
with $n_B$ again. Anti-protons annihilate 
with protons before proto-galaxies begin to cool. 
Thus, even for $Y_B \simlt 10^{-18}$, the anthropic bound 
(\ref{eq:bound-1}) is still valid.\footnote{We assume that 
$e^+$-$e^-$ and $p$-$\bar{p}$ pair annihilation emit photons 
and relativistic pions that freely escape from the gravitational 
potential of proto-galaxies.}

The anthropic bound (\ref{eq:bound-1}) has to be satisfied by 
the actual density perturbation $Q$. Assuming that $Q$ follows 
a Gaussian distribution characterized by its standard deviation 
$\sigma$, we obtain the anthropic factor 
\begin{equation}
 {\cal A}(\sigma,Y_B) \approx 
     \int_{(Q/2\times 10^{-5})> (10^{-20}/Y_B)^{6/5}} 
         dQ  e^{-(Q/\sigma)^2}
     \approx e^{-(10^{-20}/Y_B)^{12/5}(2\times 10^{-5}/\sigma)^2}.
\label{eq:exp}
\end{equation}
Since $Y_B \propto T_R^7$ for $T_R < M_1$ in the thermal 
leptogenesis scenario, this anthropic factor indeed depends 
exponentially on $T_R(m_\phi)$. 

In order to see whether this exponentially small anthropic factor
is small enough to contain the exponentially large volume factor, 
we need to be a little more specific about the inflationary landscape.
We approximate the inflationary landscape by an ensemble of 
chaotic inflation potentials, with only one parameter $m_\phi$ to be scanned.
The volume factor is given by ${\cal V} \approx e^{3N_e}$ with 
$N_e(m_\phi) \approx M_{\rm pl}/m_{\phi}$, because the largest 
value of the inflaton field for classical slow-roll inflation is 
$\phi \approx M_{\rm pl}^{3/2}/m_{\phi}^{1/2}$. 
The amount of inflation increases as $m_{\phi}$ decreases, and thus 
the volume factor can, in principle, be countered by the threshold 
behaviour and the anthropic factor we have discussed. 
The asymptotic behaviour of the anthropic factor 
${\cal A} \approx e^{-F(T_R(m_\phi))}$ is given by 
\begin{equation}
 F(T_R(m_\phi)) \propto Y_B^{-12/5}\sigma^{-2} 
   \propto T_R^{-84/5}m_{\phi}^{-2}
   \propto m_\phi^{-27.2},
\end{equation}
where $\sigma \sim m_{\phi}/M_{\rm pl}$ for chaotic inflation and 
$T_R \propto \sqrt{\Gamma(\phi)} \propto m_{\phi}^{3/2}$ for 
$m_X \simlt m_\phi \ll 2m_X$ [see (\ref{eq:3body-rate})].
Thus, as $m_\phi$ decreases, $F(T_R(m_\phi))$ increases much 
faster than $N_e(m_\phi)$ and we conclude that the exponential anthropic 
factor from thermal leptogenesis is able 
to tame the volume factor from chaotic inflation.

The anthropic factor (\ref{eq:exp}) should begin to decrease 
within the range of $T_- < T_R < T_+$ covered by the threshold 
behaviour, so that the peak of the probability distribution 
is determined by the argument in the previous section.
The requirement (\ref{eq:atTmin}) on $T_-$ now reads 
\begin{equation}
 Y_B(T_-) \simlt 10^{-20} \times (m_{\phi}/M_{\rm pl})^{5/12} 
  \approx 10^{-22},
\end{equation}
where the COBE normalization was used to determine the inflaton mass 
$m_\phi\simeq 10^{13}\; \GEV$ in chaotic inflation. This is translated
into $T_-/M_1 \simlt 10^{-1.6}$, or equivalently, 
$(T_-/T_+) \simlt 10^{-1.6} \bar{N}_e^{-1/4} (M_1/T_{R,0})\simlt 
10^{-3}(M_1/T_{R,0})$, 
with $\bar{N}_e \sim [2.4 \times 10^{18} \; \GEV / 10^{13}\; \GEV]$. 
Thus, the coupling constants for the reheating 
have to satisfy only a mild condition: 
${\rm max}(\lambda,\lambda') \simlt 10^{-2} (M_1/T_{R,0})$. 
Therefore, we obtain the upper bound on the reheating temperature 
as outlined in the previous section:
\begin{equation}
 T_{R,0} \simlt \sqrt{10^{12} \; \GEV M_1},
\label{eq:thermal-TR}
\end{equation}
and the lightest right-handed neutrino has to be somewhat lighter 
than $10^{12}$ GeV.

We have seen in this section that thermal leptogenesis 
can solve the runaway problem of chaotic inflation,  
with only a mild constraint on the coupling constants involved; 
$\lambda, \lambda'\simlt 10^{-2}$.
We have also obtained some predictions; 
the upper bound on the lightest right-handed neutrino mass 
$M_1 < 10^{12}$ GeV, and the upper bound on the reheating temperature 
of our universe (\ref{eq:thermal-TR}).
Although these predictions are not directly testable in low-energy 
experiments, they provide extra constraints on the parameter space 
of thermal leptogenesis.

\section{Non-thermal Leptogenesis}
\label{sec:neutrino}

The key ingredient in section 2 is the existence of a heavy 
particle $\psi_X$, with mass $m_X \sim 10^{13}$ GeV that is not 
scanned in a landscape. This hypothetical particle $\psi_X$
realized the sharp threshold behaviour in the reheating 
temperature, and stopped the inflaton mass runaway at $2m_X$.
Can $\psi_X$ be related to other observable physics?

The tiny masses of left-handed neutrinos combined 
with the see-saw \cite{see-saw} mechanism suggest that right-handed 
neutrinos exist, with masses of order $10^{15}$ GeV 
for order 1 Dirac Yukawa couplings and $10^{13}$ GeV 
for order $10^{-1}$ Dirac Yukawa couplings. 
Thus, a natural question arises: can a right-handed neutrino 
be identified with the hypothetical particle $\psi_X$? If this idea works, one may 
be able to make further arguments in the future why $m_X$ is 
not scanned in landscapes by looking at how the masses of right-handed
neutrinos arise. In this section, we show that one of the right-handed 
neutrinos can indeed be identified with the threshold particle.
This means that the inflaton mass, and consequently the density 
perturbation, is determined by the physics of right-handed neutrinos.

Let us assume that the inflaton decays to a pair of right-handed 
neutrinos $N_1$ with a branching ratio very close to 1 
via the interaction
\begin{equation}
 {\cal L} = \frac{1}{2}\lambda \phi N_1 N_1 + {\rm h.c.}.
\end{equation}
The right-handed neutrinos decay to a lepton $l$ and a Higgs scalar 
$h_u$ through 
\begin{equation}
 {\cal L} = \lambda'_{i \alpha} N_i l_\alpha h_u + {\rm h.c.}, 
\end{equation}
These two operators are the same as (\ref{eq:decay1}) and 
(\ref{eq:decay2}), allowing exactly the same behaviour as in 
section \ref{sec:exp}. The inflaton mass is predicted to be 
$\simeq 2m_{N_1}$, and the reheating 
temperature of our sub-universe is suppressed by roughly 
$1/N_e^{1/4}$ relative to the reheating temperature 
naturally expected when the inflaton mass is safely 
above the threshold $2m_{N_1}$.

An important difference from the scenario in the previous section, 
however, is that the baryon asymmetry does not decrease as fast as in 
(\ref{eq:7th-power}). The lepton asymmetry from the thermal 
leptogenesis certainly goes down as in (\ref{eq:7th-power}), but 
there is an extra contribution from the the leptogenesis 
in the decay of right-handed neutrinos produced by inflaton decay 
\cite{BPY}. Its lepton asymmetry is given by 
\begin{equation}
 Y_B \simeq 0.5 \frac{T_R}{m_\phi}\epsilon_{\rm CP} \equiv  
        \tilde{Y}_{B,0} \left(\frac{T_R}{T_{R,0}}\right),
\end{equation}
and decreases very slowly as $T_R$ decreases. Thus, the lepton 
(or baryon) asymmetry is dominated by this contribution for 
low reheating temperatures. Since the baryon asymmetry still depends 
on $T_R$, the sharp behaviour of $T_R(m_{\phi})$ can be used to 
solve the runaway problem. The lepton asymmetry, however, depends 
so weakly on $T_R$, as opposed to (\ref{eq:7th-power}), that a 
baryon asymmetry as low as $Y_B \sim 10^{-22}$ is not achieved 
for $T_R$ larger than the electroweak scale. 
For reheating temperatures lower than the electroweak scale, 
the baryon asymmetry decreases much faster, i.e., 
\begin{equation}
 Y_B \approx \tilde{Y}_{B,0} 
       \left(\frac{100 \; \GEV}{T_{R,0}}\right)
       \left(\frac{T_R}{100 \; \GEV}\right)^5,
\label{eq:5th-power}
\end{equation}
because the sphaleron process stops after the temperature of the 
thermal plasma drops below 100 GeV, and the baryon asymmetry 
generated at high temperature is diluted by radiation produced from inflaton decay.
Thus, the reheating temperature below the threshold, $T_-$, is 
likely to be lower than the electroweak scale, but not much less.

The anthropic bound (\ref{eq:bound-1}) on $Y_B$ (and on the density 
perturbation $Q$) was based on an assumption that the dark matter 
energy density does not change, but, with a reheating temperature 
lower than the electroweak scale, this assumption may not hold. 
If cold dark matter is a thermal relic of 
a stable particle with mass of order 100 GeV (WIMP dark matter), 
production after reheating may be highly suppressed.
The relic number density is frozen when the plasma temperature is 
around 10 GeV, and is diluted by a factor of order $(T_R/10\; \GEV)^5$ 
due to the following entropy production from inflaton decay. 
Gravitinos produced from thermal scattering are also a candidate for  
dark matter in gauge-mediated supersymmetry breaking scenarios, and 
their number density also decreases as the reheating temperature 
decreases---not exactly in the same manner for $T_R$ above the 
electroweak scale, though. 

For $T_R$ sufficiently less than 10 GeV for WIMP dark matter, 
the dominant component of dark matter is left-handed neutrinos.
The effect of a lower dark matter density is incorporated by taking 
$T_{eq}$ to be $m_{\nu} \approx 0.03 \; \EV$ in 
(\ref{eq:rho})--(\ref{eq:n-B}). Thus, the anthropic bound 
(\ref{eq:bound-1}) is replaced by 
\begin{equation}
 \left(\frac{10^{-21}}{Y_B}\right)^3 
 \left(\frac{2\times 10^{-5}}{Q}\right)^{2.5} < 1.
\label{eq:bound-2}
\end{equation}
Note that the galactic comoving scale enclosing a baryonic mass 
of $M_{\rm gal}$ is longer than the free-streaming length 
of neutrinos for $Y_B$ as low as $10^{-21}$, and the neutrinos 
can be treated as cold dark matter 
in the earlier stage of galaxy formation. The same argument as in the 
previous section leads to an exponential cut off for low reheating 
temperature and small inflaton mass, and it is sufficient to take  
$T_-$ so that 
\begin{equation}
 Y_B(T_-) \simlt 10^{-21} \times (m_{\phi}/M_{\rm pl})^{5/12} 
 \approx 10^{-24}.
\label{eq:bound-2'}
\end{equation}
Thus, $T_- \sim 1 \; \GEV$ is sufficient for 
$T_{R,0} \sim 10^{5.5}\; \GEV$, and $T_- \sim 10 \; \GEV$ for 
$T_{R,0} \sim 10^{10.5} \; \GEV$.
Since $Y_B$ depends on $T_R$ quite sensitively in (\ref{eq:5th-power}) 
for low $T_R$, the exponential anthropic factor in (\ref{eq:exp}) 
is powerful enough to tame the volume factor of chaotic inflation.
At the same time, one will notice that various uncertainties already 
mentioned in footnotes in the previous section do not lead to a large 
uncertainty in the constraint on $T_-$. 

One can also see that the bound (\ref{eq:bound-2}) 
(and (\ref{eq:bound-2'})) is valid for gravitino dark matter as well, 
essentially because left-handed neutrinos become the dominant 
cold dark matter and the other components are small and no longer 
relevant. 

Now we can derive the upper bound on the reheating temperature 
of our universe, using the constraint (\ref{eq:bound-2'}).
Higher reheating temperature $T_{R,0}$ (and higher $T_+$) requires 
larger coupling constant, because 
$T_+ \approx$ min($\lambda$, $\lambda'_{i=1,\alpha}$)$\times 10^{15}$
GeV, where $m_\phi \sim 10^{13}$ GeV is used for chaotic inflation. 
This means larger $T_-/T_+ \sim {\rm max}(\lambda,\lambda')/(4\pi)$. 
Since the constraint (\ref{eq:bound-2'}) sets an upper limit on $T_-$, 
$T_{R,0}$ cannot be arbitrary high. 
Let us assume $\bar{N_e}^{1/4} \sim 10$. 
For $T_{R,0}\sim 10^7$ GeV, $T_+ \sim 10^8$ GeV, which means that 
${\rm min}(\lambda,\lambda'_{i=1,\alpha}) \sim 10^{-7}$. The threshold 
behaviour of $T_R(m_\phi)$ can cover the range from $T_+$ to 
$T_- \sim ({\rm max}(\lambda,\lambda')/4\pi) T_+ \simgt$ 1 GeV. Thus, 
for $\lambda \sim \lambda'_{i=1,\alpha}\sim 10^{-7}$, the anthropic 
factor is already sufficiently exponentially small at the lower 
end of the threshold. For $T_{R,0} \sim 10^8$ GeV, $T_-$ cannot be 
less than 100 GeV, and the anthropic factor is not exponentially 
small. The upper bound on the reheating temperature of our 
sub-universe, $T_{R,0}$, is only slightly above $10^7$ GeV.

At the same time we obtain the upper bound on the Dirac Yukawa 
coupling constants of the lightest right-handed neutrinos:\footnote{
Dirac Yukawa coupling constants of other right-handed neutrinos are 
not constrained. Thus, the CP asymmetry can be large enough to 
account for the lepton/baryon asymmetry of order $10^{-10}$.} 
$|\lambda'_{i=1,\alpha}| \simlt 10^{-7}$. 

Leptogenesis through inflaton decay accounts for the baryon 
asymmetry of our universe 
$Y_{B,0} \sim (0.4\mbox{--}0.9)\times 10^{-10}$ 
only when $T_{R,0}\simgt 10^6$ GeV, since the CP asymmetry  
in (\ref{eq:CP}) cannot be more than order one. 
Thus, we see that the range of reheating temperatures of our universe 
is narrowed down to one order 
of magnitude, $10^6 \; \GEV \sim 10^7$ GeV.
This means that the CP phase in leptogenesis is no less than of order 
0.1. If there is no significant accidental cancellation with 
other combinations of CP phases in the lepton sector, 
there is a chance that CP violation will be seen in the 
next-generation of neutrino oscillation experiments.

The other prediction, $|\lambda'_{i=1,\alpha}| \simlt 10^{-7}$, 
implies that the mass matrix of left-handed neutrinos is virtually 
of rank 2, if there are 3 right-handed neutrinos. Some predictions 
on the rate of neutrinoless double beta decay are available in the 
rank 2 case; see \cite{FHY} for more information. 

These predictions are based on chaotic inflation, where the inflaton 
mass is determined by the COBE normalization 
$m_{\phi}\simeq 10^{13}$ GeV. We immediately learn that 
one of the right-handed neutrinos has a mass $m_\phi/2$.

With weak-scale supersymmetry breaking\footnote{
In supersymmetric theories, the Affleck--Dine mechanism is another way 
to generate the baryon asymmetry. However, a negative mass squared is 
generated only when the K\"{a}hler potential satisfies certain 
conditions. If a landscape supports a region where the conditions 
are satisfied, an extra consideration is necessary.} arising 
from gauge mediation, our prediction for the range of 
reheating temperatures $T_{R,0}$ has a significant further implication.
In these theories, the gravitino is the best dark matter 
candidate for a wide range of gravitino masses, 
$100 \; \KEV\mbox{--}100$ GeV. The relic density of gravitinos 
produced from the thermal plasma is given by \cite{BBB}
\begin{equation}
\Omega_{3/2}h^2 \simeq 0.2 \left(\frac{T_R}{10^6\; \GEV}\right)
  \left(\frac{10 \;\MEV}{m_{3/2}}\right)
  \left(\frac{m_{\tilde{g}}(100\; \GEV)}{1\; \TEV}\right)^2, 
\end{equation}
and depends on two totally unknown parameters $m_{3/2}$ and $T_R$.
Only the ratio $(T_R/m_{3/2})$ can be determined from 
the observed value of $\Omega h^2 \simeq 0.113$.
But, now that the reheating temperature is essentially predicted
the gravitino mass is determined, $m_{3/2}\simeq$ 20 MeV--200 MeV. 
This in turn determines the fundamental scale of supersymmetry 
breaking, $\sqrt{F} \approx 10^8\mbox{--}10^9$ GeV.

\section{Conclusions and Discussion}
\label{sec:concl}

If the fundamental theory allows more than one vacuum, and if many
vacua are realized cosmologically, then cosmology may 
play an important role in choosing the low energy theory and its
parameters. In inflationary landscapes, where models and parameters 
of inflation are scanned cosmologically, sub-universes with a small 
inflaton mass dominate the volume of the universe, making slow-roll 
inflation a probable consequence. The problem of inflationary 
landscapes \cite{FHW,GV}, however, is that the probability 
distribution tends to be exponentially sensitive to some late-time 
cosmological parameters, and hence does not correctly predict their 
observed values. We call this the inflation runaway problem.

Landscapes and many-universes are about the world outside the 
horizon of our universe, and we usually think that no sign of them 
can be seen. One also usually considers that the anthropic conditions 
depend on only a limited number of relevant parameters, most of which 
have already been measured, so that there are no predictions. 
Without a well-understood top-down tool to figure out the statistic 
distribution of vacua on a landscape, it seems almost impossible 
to determine theories or parameters that are yet to be tested in experiments.
It is important to note, however, that inflationary landscapes 
have a generic prediction: the probability distribution is 
exponentially sensitive to some parameters. This is where the runaway 
problem comes from. The problem is so severe that there may be very
few possible solutions, from which indications may be obtained for 
particle physics to be explored in the future. 

Some ideas have already been proposed to evade this problem. 
For instance, eternal inflation of false-vacuum type or chaotic-type 
may strongly select only one inflationary region in a landscape 
\cite{FHW,GV}. The observational consequences for the curvature of the 
universe and for lower multipoles of CMB spectrum are discussed in 
\cite{Susskind05} and references therein.
Maybe the fundamental theory of phsycis has an inflationary landscape 
carefully designed to avoid the volume factor exponentially sensitive 
to some parameters \cite{FHW,Linde-curvaton}. If such landscapes 
involve chaotic type inflation, we may expect tensor perturbations. 
A fundamental cut off scale parametrically larger than the observed 
Planck scale may also have something to do with the apparent fine 
tuning problems of certain low-energy supersymmetric theories 
\cite{IIY}.
An alternative is to have density perturbations generated by 
a curvaton or by modulated reheating \cite{FHW,GV}, whose observational
consequences are discussed in the literature 
\cite{curvaton,nonGauss,Linde-curvaton}.

In this article, we have proposed a new solution to the runaway 
problem, which is connected to the {\em particle physics} of 
inflaton decay and baryogenesis. 
In spite of the exponential distribution from the inflationary 
landscapes, observables are predicted successfully in the middle 
of the allowed anthropic windows. 
The key ingredient is to distinguish between parameters describing 
inflation and those that are directly observed. 
The latter are not necessarily given by simple power-law functions 
of the former. For instance, the reheating temperature drops 
sharply as the inflaton mass approaches the threshold  
for two body decay. In this case, the 
inflationary landscape predicts the inflaton mass to be close 
to twice the mass of the threshold particle. 

A sufficiently sharp threshold behaviour for $T_R$ requires small 
couplings in operators relevant to reheating, and hence a low $T_R$ 
in our universe that depends on the cosmological scenario: 
$T_R \simlt 10^{12}$ GeV in the thermal leptogenesis scenario of 
section \ref{sec:exp}, and $10^6 \; \GEV \simlt T_R \simlt 10^7 \; 
\GEV$ in the inflaton-decay leptogenesis scenario of section
\ref{sec:neutrino}. Chaotic inflation is assumed. 
In both sections, the anthropic limit was set by
requiring that bremsstrahlung and runaway hierarchical cooling 
of proto-galaxies start before protons decay. Although we have ignored
various uncertainties,
the upper bound obtained on $T_R$ illustrates a completely new way 
to obtain information on particle cosmology from anthropic reasoning.

In section \ref{sec:neutrino} the particle produced by inflaton decay 
was identified with a right-handed neutrino, allowing 
the inflaton mass, and some other parameters including 
the normalization of density perturbations, to be set by physics 
of the neutrino sector. This provides a mechanism for understanding 
the values of inflation parameters in terms of observable 
particle physics. In this particular scenario, there are several
predictions that again demonstrate how this line of anthropic argument
can lead to observable tests. One CP violating phase of the neutrino
sector must be of order 0.1 or larger. 
The low energy neutrino mass matrix is forced to be essentially rank 2, so that the
rate for neutrinoless double beta decay will be constrained 
by a particular relationship 
involving the neutrino mixing angles and the observed neutrino masses. 
Furthermore, in supersymmetric
theories with gauge mediation, dark matter is predicted to be
gravitinos with mass in the range 20--200 MeV. This corresponds to a
fundamental scale of supersymmetry breaking of $10^8$--$10^9$ GeV,
which can be tested by precise measurements of the superpartner spectrum
and by the nature of the decay of the next-to-lightest superpartner 
\cite{DESY}.

It should be noted that the above predictions are based on an 
assumption that the anthropic limit is set by the comparison between 
the rate of bremsstrahlung cooling and proton decay. When various 
other anthropic conditions are imposed, some of which are listed 
in footnotes in section \ref{sec:exp}, the anthropic limit in the 
$Y_B$--$Q$ plane may be stronger, and the upper bounds on $T_R$ may 
become weaker. Including other uncertainties in galaxy 
formation and in the anthropic conditions, 
the above predictions may be modified. 
There is no doubt that further studies in astrophysics will be important 
in making the predictions more precise and reliable. As a step in 
this direction, we discuss possible effects of particle evaporation 
from proto-galaxies in the appendix, concluding 
that the upper bound on $T_R$ will not be greatly changed, partly 
because,  in scenarios in section 
\ref{sec:exp} and \ref{sec:neutrino},
$Y_B$ is highly sensitive to $T_R$.
 
As the astrophysics and anthropic conditions become better understood, 
the upper bounds of the reheating temperature will change 
numerically. But, the idea for stopping the runaway by a combination 
of threshold effects and exponential anthropic factors will survive; 
extra anthropic conditions  would only strengthen our 
conclusion that the exponential distribution of inflationary 
landscapes can be tamed. No matter how sophisticated and complicated 
the astrophysical analysis becomes, the upper bound on the 
reheating temeperature persists as long as the threshold behaviour 
of the inflaton decay plays a crucial role.
 
\section*{Acknowledgements} 

This work was supported in part by the Director, Office of Science, 
Office of High Energy and Nuclear Physics, of the US Department of 
Energy under Contract DE-AC03-76SF00098 and DE-FG03-91ER-40676, 
DE-AC02-05CH11231, in part by the National Science Foundation 
under grant PHY-00-98840 (L.H., T.W.), and in part by a Humboldt 
Research Award (T.T.Y.). 
We thank B.~Feldstein and K.-I.~Izawa for discussion.
T.W. thanks the Miller Institute for Basic Research in Science.
T.T.Y. thanks the DESY theory group for hospitality, where a part 
of this work was done.

\appendix 

\section*{Appendix: Evaporation}

In proto-galaxies that cannot cool, some fraction of the baryons 
in the high-energy tail of the Boltzmann distribution  
have sufficient speed to escape, as noted in 
\cite{TARW}. We ignored this effect in the main text, and devote this
appendix to investigating the extent to which the conclusions could be 
changed by evaporation. 

There are two important points in discussing evaporation from 
proto-galaxies. One is that the evaporation of particles from 
isolated proto-galaxies results in their gravitational contraction 
\cite{evap-contr}.  The other is that the proto-galaxies are 
comprized of multiple components: dark matter particles and 
baryons (including electrons and possibly their anti-particles).
We begin for simplicity with the evolution of proto-galaxies 
made purely of cold dark matter particles, and later extend 
the discussion to take account of baryons. 

Nearly 1\% of particles in the Maxwell--Boltzmann distribution 
have velocities larger than the escape velocity of proto-galaxies 
and hence evaporate. The time scale of evaporation is set by how 
quickly the lost population of particles in the high-energy tail of the
Boltzmann distribution is restored by particle--particle interactions.
Two-body interactions via gravity always exist, no matter what the 
nature of the dark matter particle, with an efficiency given by 
\begin{equation}
 \Gamma_{grav.} \approx \frac{(G_N \rho)^{1/2}}{N_{DM}},
\end{equation}
where $N_{DM}$ is the typical number of dark matter particles 
in one proto-galaxy \cite{evap-contr}. Note that the number of baryons 
is very large, of order $N_B \sim 10^{68}$ for $M_{\rm gal}\simeq 
10^{11} M_{\odot}$, and $N_{DM}$ is extremely large if dark matter 
is composed of elementary particles with a moderate mass, such as 100 GeV, or even 
lighter. Thus, the gravitational encounter of two dark-matter particles
is not effective within the time scale of bremsstrahlung cooling.

When dark-matter particles have two-body non-gravitational scattering 
with a cross section $\sigma_{DD}$, the relaxation due to this 
scattering occurs at the rate $n_{DM} \sigma_{DD} v_{\rm vir}$, 
where $n_{DM}$ is the number density of dark-matter particles.
The evaporation is roughly 1\% of the relaxation rate, so that 
evaporation of dark-matter particle does not occur 
before bremsstrahlung cooling, if 
\begin{equation}
 10^{-2} n_{\rm DM} \sigma_{DD} v_{\rm vir} \simlt \Gamma_{\rm Brems}. 
\label{eq:relax-brem}
\end{equation}
Using (\ref{eq:rho}) and (\ref{eq:brems}), this holds true as long as
\begin{equation}
 \sigma_{DD} \simlt 10 \; {\rm pb} 
   \left(\frac{m_{DM}}{100 \; \GEV}\right)
   \left(\frac{Y_B}{10^{-22}}\right)
   \left(\frac{2\times 10^{-3}}{Q}\right)
\label{eq:DD-sigma}
\end{equation}
for dark matter with $\rho_{DM}/s \simeq 0.8$ eV, and 
\begin{equation}
 \sigma_{DD} \simlt 10^{-22.5} \; {\rm b} 
    \left(\frac{Y_B}{10^{-24}}\right)
    \left(\frac{2\times 10^{-3}}{Q}\right) 
\label{eq:nn-sigma}
\end{equation} 
for neutrino dark matter with $n_\nu/s \simeq 10^{-1.5}$.
It is clear that cross sections for axion--axion scattering and 
neutrino--neutrino scattering satisfy (\ref{eq:DD-sigma}) and 
(\ref{eq:nn-sigma}), respectively, for the values of $Y_B$ given in 
the right-hand sides.
Thus, the anthropic bounds (\ref{eq:bound-1}) and (\ref{eq:bound-2}) 
in section \ref{sec:exp} and \ref{sec:neutrino} are unaffected by the 
evaporation of dark matter in such cases.

While the inequality (\ref{eq:DD-sigma}) is satisfied by most dark 
matter candidates, for example WIMPs, it could be violated when the 
interactions between dark matter particles are significant. 
In that case, the anthropic bound (\ref{eq:bound-1}) in section 
\ref{sec:exp} will be affected; dark-matter evaporation leads 
to the gravitational contraction of proto-galaxies, and then the 
baryon number density is also enhanced and bremsstrahlung cooling 
begins to work. At some point in the gravitational contraction due 
to evaporation and quasi-static bremsstrahlung cooling, 
runaway cooling starts.
Thus, evaporation changes the estimate of the time scale 
before the hierarchical cooling starts, 
which is to be compared with the proton lifetime in (\ref{eq:proton}). 
However, even for such dark matter candidates, the estimated anthropic 
bound in (\ref{eq:bound-1}) is not 
greatly affected by the evaporation of dark matter particles, 
since unitarity implies that(\ref{eq:DD-sigma}) is not violated 
by a large amount---not by of order $10^{10}$.
When the anthropic bounds on $Y_B$ is translated to an upper bound 
on the reheating temperature $T_R$, the effect will be quite minor 
since $Y_B$ is very sensitive to $T_R$ in (\ref{eq:7th-power}).

Let us now bring baryons into the discussion, ignoring the evaporation of 
dark matter particles. Baryons-baryon scattering has a large cross 
section and the Boltzmann distribution is 
repopulated much more quickly. 
We do not go into a detailed estimate of the evaporation rate of 
baryons, but an important point is that this evaporation is 
accompanied by a gravitational contraction of the baryons. 
As long as there is no 
efficient energy transfer between dark-matter particles and baryons, 
baryons fall into the potential well of proto-galaxies as the 
evaporation proceeds, so that the evaporation soon stops. 
Thus, the discussion in the main text is unaffected by the evaporation 
of baryons (and electrons) if the energy transfer between dark matter 
particles and baryons does not occur effeciently.

Consider the case that the dark-matter particle is lighter than the 
proton, for example the case of axions or neutrinos. Since the 
axion-nucleon and neutrino-nucleon scattering cross 
sections are so small, an inequality similar to (\ref{eq:nn-sigma}) is safely 
satisfied. Thus, the energy transfer does not take place between 
axion or neutrino dark matter with baryons before bremsstrahlung 
cooling starts, and the conclusion of the previous paragraph holds.

When dark matter particle has a mass of order $100$ GeV--$1$ TeV, 
the kinetic energy of a proton typically increases by of order unity in a 
single collision with a dark-matter particle. Once a proton is 
kicked by that amount, it has a good chance to escape the 
gravitational potential. The typical rate for this to happen is 
$n_{DM} \sigma_{DN} v_{\rm vir}$, which is to be compared with 
$\Gamma_{\rm Brems}$. Energy transfer from dark matter to baryons 
does not occur before bremsstrahlung cooling, if the 
dark-matter nucleon scattering cross section $\sigma_{DN}$ satisfies 
\begin{equation}
 \sigma_{DN} \simlt 0.1 \; {\rm pb} 
   \left(\frac{m_{DM}}{100 \; \GEV}\right)
   \left(\frac{Y_B}{10^{-22}}\right)
   \left(\frac{2\times 10^{-3}}{Q}\right).
\label{eq:DN-sigma}
\end{equation}
Most candidates for dark matter have a scattering cross section with 
protons that is much smaller than 0.1 pb, and hence the discussion 
in section \ref{sec:exp} is not affected by baryon evaporation 
from proto-galaxies. 

In the rare cases that energy is transferred from dark matter to 
baryons, the baryons may evaporate efficiently from proto-galaxies 
since dark matter particles keep supplying energy to the baryons 
allowing them to escape. 
Since protons and electrons are light and minor components in 
proto-galaxies, the energy supply continues 
until proto-galaxies loose virtually all their protons and electrons. 
Thus, bremsstrahlung cooling should 
work in a time scale shorter than that of the energy transfer to 
baryons and evaporation of baryons. This anthropic condition sets 
a constraint in the $Y_B$--$Q$ plane in addition to the proton lifetime 
constraint (\ref{eq:proton}).
An argument can be constructed essentially in the same way as in the main 
text, and in the end, the main qualitative result is maintained even 
with evaporation: namely, an exponential anthropic factor results
from the failure of structure formation, and an upper bound on the reheating 
temperature is required to make $Y_B(T_-)$ small enough. 
Since (\ref{eq:DN-sigma}) is not a stringent constraint, and since 
$Y_B$ is very sensitive to $T_R$, the upper bound on $T_R$ is not 
affected very much, even quantitatively.

It should be noted, however, that the discussion so far, both in the 
appendix and hence in the main text, assumes that proto-galaxies are 
homogeneous, allowing a crude estimate of the impact of 
evaporation. The limit on the cross section, such as 10 pb, may change
a little when one takes into account the concentration of particles 
at the centre of the proto-galaxies and the profiles of the matter distribution 
and gravitational potential. The limit may also change when 
additional anthropic conditions, such as those in footnotes in 
section \ref{sec:exp}, are taken into account, or when 
$M_{\rm gal} \sim 10^{11} M_{\odot}$ is replaced by an anthropic 
lower bound on $M_{\rm gal}$, yet to be determined precisely. 
Further study of such issues may change the argument in this appendix 
quantitatively, but the way of thinking will remain valid.

\end{document}